\documentclass[%
 reprint,
 amsmath,amssymb,
 aps,
]{revtex4}

\usepackage{graphicx}
\usepackage{dcolumn}
\usepackage{bm}
\usepackage{xcolor}
\usepackage{hyperref}
\usepackage{url}

\newcommand{\bea}{\begin{eqnarray}}
\newcommand{\eea}{\end{eqnarray}}

\begin{document}


\title{Real time evolution of scalar fields with kernelled Complex Langevin equation}

\author{Nina Maria Lampl}
\author{D\'enes Sexty}
\affiliation{%
Institute of Physics, NAWI Graz, University of Graz,\\
Universitätsplatz 5, Graz, Austria }%



\date{\today}

\begin{abstract}
The real time evolution of a scalar field in 0+1 dimensions is investigated on a complex time contour. The path integral formulation of the system has a sign problem, which is circumvented using the Complex Langevin equation. Measurement of the boundary terms allow for the detection of correct results (for contours with small real time extents) or incorrect results (at large real time extents), as confirmed by comparison to exact results calculated using diagonalization of the Hamiltonian. We introduce a constant matrix kernel in the Complex Langevin equation, which is optimized with the requirement that distributions of the fields on the complexified manifold remain close to the real manifold. We observe that reachable real times are roughly twice as large with the optimal kernel. We also investigate field dependent kernels represented by a neural network for a toy model as well as for the scalar field, providing promising first results.
\end{abstract}

\maketitle


\section{Introduction}

Ab initio calculations of real-time quantities in Quantum field theories (QFT)
is one of the main challenges of theoretical physics.
Non-equilibrium physics of quantum field theories can be tackled using various approximations such as the 
classical approximation, 
functional equations 
and the kinetic approximation, 
however these approaches have systematic errors which are very hard to control.
In contrast, lattice discretisation has relatively easy to 
control systematics, and has proven to be a very useful tool in studying QFTs.
Lattice calculations 
are usually formulated in imaginary time, leading to a theory
suitable for importance sampling Monte Carlo simulations. 
In contrast, using real-time lattices (e.g. with the aim of investigating non-equilibrium phenomena)
leads to a path integral with complex measure $\exp(i S)$ with the (real valued) action $S$, so naive importance sampling 
simulations are invalidated by the sign-problem.
Even some equilibrium quantities are affected by this problem: real-time correlators
such as those needed in the Kubo formulas for hydrodynamical transport coefficients,
spectral functions of bounded states or parton distribution functions,
while in principle accessible starting
from the euclidean theory, they require the inversion of an  
integral kernel, which often leads to 
an underdetermined system of equations, usually solved with 
Bayesian methods or with the usage of further assumptions or modeling of the system 
(see the recent review \cite{Rothkopf:2022ctl}).
Other problems, such as the tunneling rate between metastable vacua in QFTs
or the equilibration of the matter in the initial stages of a heavy ion collision
present a non-equilibrium
problem inaccessible to euclidean simulations.

The sign problem can be (partially) circumvented in various ways,
for a recent reviews of different methods see \cite{Berger:2019odf,Alexandru:2020wrj,Nagata:2021ugx}.
In this paper we concentrate on the Complex Langevin equation (CLE)\cite{Klauder:1983nn,Parisi:1984cs},
which was proposed to circumvent
the sign problem by complexifying the simulation manifold, such that the original theory is recovered
in expectation values using analytic continuation of the observables. 
Among many applications ranging from systems 
at nonzero density or theta-term to condensed matter models with fermionic imbalance  (for a recent review see \cite{Berger:2019odf,Attanasio:2020spv}),
the CLE was also used to study Quantum Field theories in Minkowski spacetime \cite{Huffel:1984mq,Callaway:1985vz,Nakazato:1985zj}.
The CLE was first applied to investigate non-equilibrium physics in \cite{Berges2005b}. In 
\cite{Berges:2006xc,Berges:2007nr} a variant of the Schwinger-Keldysh contour was introduced for the 
lattice discretisation, which allows equilibrium as well as non-equilibrium simulations, and 
a scalar oscillator and a 3+1 dimensional SU(2) pure gauge theory was investigated. 

The problem of real-time simulations has also been tackled by 
a close relative of the Complex Langevin method:
optimized complexified manifolds \cite{Alexandru:2016gsd,Alexandru:2020wrj}
(as inspired by simulations on Lefschetz-thimbles \cite{Cristoforetti:2012su}), observing that similarly to complex Langevin,
real time extents with the magnitude of about a period of the oscillator are reachable (for larger time extents the costs of the simulation get too high).

The Langevin equation (as well as the CLE) allows a modification with a kernel \cite{Soderberg:1987pd,Okamoto:1988ru,Okano:1991tz}, which leaves 
the stationary solution of the corresponding Fokker-Planck equation unchanged.
In this paper we introduce a kernel in the Complex Langevin equation such that
boundary terms are reduced. We search for an optimal kernel using Machine Learning inspired
methods. We first use field independent kernels, afterwards we do the first exploratory steps in finding
field dependent kernels. We study an anharmonic quantum oscillator 
as well as a toy model to help understand the behavior of the kernel optimization process.
A study with a similar aim has appeared recently \cite{Alvestad:2022abf} which was carried out
independently from our study. Another recent application of an anisotropic, constant kernel has been 
studied in real-time simulations of a SU(2) pure gauge theory \cite{Boguslavski:2022dee}.

The paper is organized as follows: In section \ref{sec2}, we introduce the Complex Langevin equation,
boundary terms, kernels. 
Section \ref{realtimesection} is devoted to the anharmonic quantum oscillator:  we show results using naive CLE, 
discuss our optimization procedure for a constant kernel, and show the results of 
the optimally kernelled CLE. In section \ref{secml} we discuss field dependent kernels, described
with a suitable ansatz as well as with a neural network, as applied to a toy model and the quantum oscillator.
Finally in section \ref{secconc} we summarize our results and conclude.

\section{Complex Langevin  eq., Kernels  and numerical setup}
\label{sec2}

In the path integral formulation, a quantum mechanical system is defined in terms of its 
path integral with the measure $ \exp(iS) $ with the action $S$.
For an imaginary time contour the action is imaginary, giving a real and positive measure.
For a non-strictly imaginary time-contour
the action is in general complex, which leads to the appearance of the sign-problem, invalidating 
importance sampling Monte Carlo simulations. To circumvent this problem, we 
use the complex Langevin equation (CLE) \cite{Parisi:1984cs,Klauder:1983nn}, which is given by
\bea
{  \partial \phi_i(\tau) \over \partial \tau} = - { \partial S \over \partial \phi_i } + \eta_i(\tau),
\eea
where we introduced the Langevin time $ \tau$, and $i$ indexes the fields on a space-time lattice 
along some complex-time contour,
and $ \eta_i(\tau)$ is a Gaussian noise term satisfying $ \langle \eta_i (\tau) \rangle =0 , \ \ 
\langle \eta_i (\tau) \eta_j (\tau') \rangle = 2 \delta_{ij} \delta(\tau-\tau')$.

The complex Langevin equation is known to sometimes produce wrong results 
due to either insufficient decay of the distribution at infinity or non-holomorphic actions \cite{Aarts:2009uq,Aarts:2011ax,Aarts:2017vrv,Seiler:2023kes}.
These can be characterised by boundary terms, which can be measured using the complexified process only
\cite{boundaryterms1,boundaryterms2}.
The boundary terms are given in terms of some norm $N(\phi)$ and a cutoff variable $Y$, such that $N(\phi)<Y$
defines a compact set containing the original real manifold. To extract the boundary terms for some observable $ O(\phi)$, one has to measure
\bea
  \lim_{Y \rightarrow \infty } \left\langle \theta( Y - N(\phi) ) \sum_i ( \partial_i + K_i ) \partial_i O(\phi) \right\rangle,
\eea
such that a nonzero value signals an incorrect CLE result. (For a more detailed explanation, see \cite{boundaryterms1,boundaryterms2}).

Setting up the Langevin equation one has a great amount of freedom. 
One of the allowed modifications is called a kernel \cite{Soderberg:1987pd,Okamoto:1988ru,Okano:1991tz}. 
The kernelled Langevin equation for a single degree of freedom reads
\begin{eqnarray} \label{singleKCLE}
    { d z \over d \tau} = -K(z) { \partial S \over \partial z } + { \partial K(z) \over \partial z} + \sqrt{ 2 K(z)} \eta(\tau),
\end{eqnarray}
where we have defined $K(z)$, the field dependent kernel.  For scalar fields $\phi_i,\ i=1..N$ we 
write 
\bea
\label{kernlang}
{d \phi_i \over d\tau} = 
 -H_{ij}(\phi) H^T_{jk}(\phi) {  \nabla_k S }
+ \partial_k ( H_{ij}(\phi) H^T_{jk}(\phi)) +  H_{ij}(\phi) \eta_j,
\eea
where $\nabla_i$ is differentiation with respect to the $i$-th scalar 
variable and using a field dependent matrix $H_{ij}(\phi)$, which is assumed to be 
symmetric below. In older literature the square of this matrix  
$ K_{ij}(\phi) = H_{ik}(\phi) H_{kj}(\phi) $ is called the kernel. 
Such a kernel can be used for e.g. Fourier acceleration of 
the Langevin simulations \cite{Batrouni:1985jn}.
Using 
$H_{ij}=\delta_{ij}$ gives back the original Langevin equation.
Note that a constant kernel $H_{ij}=C \delta_{ij}$ is just a redefinition of the 
Langevin time using the scaling $ \tau'= \sqrt{C} \tau $. 
 It has been shown in \cite{Aarts:2012ft} that a subclass of kernels is 
equivalent to setting up the Langevin equation after a variable change in 
the path integral has taken place. 
It's easy to show that a kernel changes the scalar
Fokker-Planck equation into 
\bea
\partial_\tau P (\phi,\tau) = \nabla_i ( H_{ik} H_{jk} ( \nabla_j + \nabla_j S )) P(\phi,\tau),
\eea
where $P(\phi,\tau)$ is the probability distribution of the variables at Langevin
time $\tau$.  From this one notices that the standard $P=e^{-S}$ remains a stationary solution 
of the Fokker-Planck operator.
The stability properties as well as the uniqueness of this solution 
might change in general.
 It is shown below that with a careful choice of the kernel one can improve 
the stability properties of the equilibrium distribution, and in practice 
one might use a kernel to reduce boundary terms in the simulation 
of the Complex Langevin equation.

\section{Real-time evolution of the anharmonic quantum oscillator}
\label{realtimesection}

The model we investigate is a scalar field theory with a quartic self-coupling, defined by the action
\bea
 S = \int dt d^d x \left( {1\over 2 } \dot \phi ^2  -{1\over 2 } (\nabla \phi)^2  - V(\phi) \right), 
 \quad V(\phi) =  {1\over 2 } m^2 \phi^2 + {\lambda \over 24} \phi^4.
\eea 
The path integral describing the quantum mechanical evolution of the system in turn contains the measure 
$ \exp ( i S ) $. 
An imaginary temporal path starting at $t=0$ and ending at $ t=-i \beta$ allows the description
of a thermal state with temperature $T = 1/\beta $, when one uses periodical boundary conditions.
To study real-time correlators (or in general, real time evolution), we discretise the system 
on a complex time-contour \cite{Berges:2006xc}, using $C_t,\ \ t=0 ... N$ with $C_0=0$ and $\textrm{Re} C_N=0$, and the complex time-step $ \Delta_t= C_{t+1}-C_t$. 
We use periodic boundary conditions $ \phi_0(x)=\phi_N(x)$ such that the system is in thermal equilibrium.
The discretised action for a $d+1$ dimensional theory then reads
\bea
S = {1 \over 2} \sum_{t,x} \left( {  (\phi_{t+1}(x) -\phi_t(x))^2 \over \Delta_t   }
 - {\Delta_t \over 2}\sum_{\nu=1}^d \left( (\phi_t(x+\hat \nu) - \phi_t(x) )^2  +  (\phi_{t+1}(x+\hat \nu) - \phi_{t+1}(x) )^2 \right)
 - \Delta_t \left( V(\phi_t) + V(\phi_{t+1}) \right) 
 \right) 
\eea
with the field $\phi_t(x) = \phi(C_t,x) $ and $ x +\hat \nu$ denoting the neighbouring lattice site in spatial direction $\nu$. 
In the following we consider a 0+1 dimensional system corresponding to an anharmonic quantum oscillator.

We have used a triangle contour on the complex time plane, starting at $ t_\textrm{start}=0$,
following a straight line to the turning point $ t_\textrm{mid}=t_\textrm{max} - 0.01 i \beta$, and again
following a straight line to the endpoint $ t_\textrm{end}=-i \beta $. 
(We have checked that our results remain unchanged when using a horizontal upper part of the
triangle contour such that $ t_\textrm{mid}=t_\textrm{max}$.)
We use periodic boundary conditions
such that the path integral describes the oscillator in thermal equilibrium.
This temporal path is discretised to $N_t/2$ equidistant points on the upper part 
of the contour and $N_t/2$ equidistant points on the lower part.
We use the parameters $ m^2=1, \lambda=24 $ and $ \beta=1$, and various $t_\textrm{max} $ values. An improved update equation as well as adaptive step size are used for all calculations. The maximal Langevin step size is chosen to be $\Delta \tau=10^{-5}$. 

Our two main observables are the equal-time two point function $ \langle \phi(t)^2\rangle $ (which is 
time independent in thermal equilibrium), and the unequal-time two point function 
$ \langle \phi(0) \phi(t) \rangle $. The exact results for these correlators 
for a general complex $t$ parameter
can be calculated using the truncation of the Hamiltonian operator in particle number basis, and performing the exponentiation of the truncated matrix in its eigenbasis.
To get precise results for the purposes of this study, one needs to truncate
to about $ N=32$ states.

Judging whether the CLE (with or without a kernel) delivers correct results can proceed in 
various ways. As the system we are studying is in thermal equilibrium,
all correlators should be time translation invariant. First, the easiest check is therefore
to study an equal-time two point function (The one point function is zero, even if the CLE results are wrong for other observables.),
 $ \langle \phi(t)^2 \rangle$. Significant deviations from a constant in time
 mean that boundary terms are non-zero and the results are incorrect.
Second, we can explicitly measure boundary terms as discussed above. Third, for this 
simple model we can calculate the exact results using diagonalization, this is 
however going to be infeasible for eventual applications of this method
for scalar field theories as we study the theory in $d>0$ spatial dimensions.

\subsection{Boundary terms}

\begin{figure}[h]
\begin{center}
    \includegraphics[width=0.32\columnwidth]{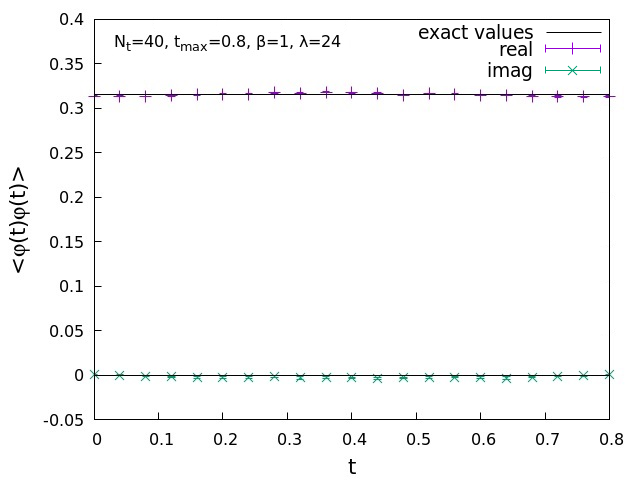}
    \includegraphics[width=0.32\columnwidth]{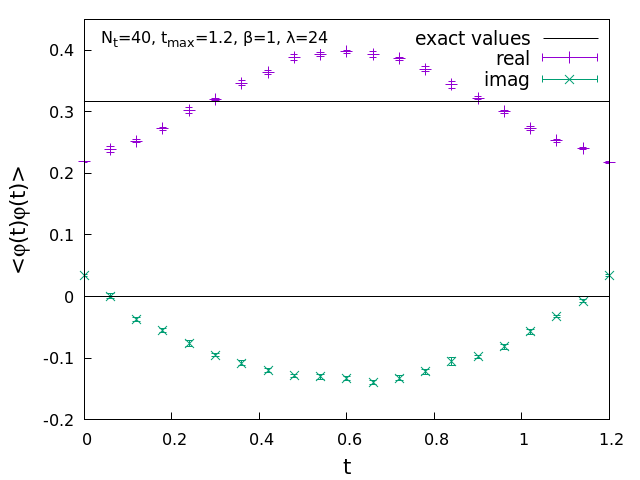}
    \includegraphics[width=0.32\columnwidth]{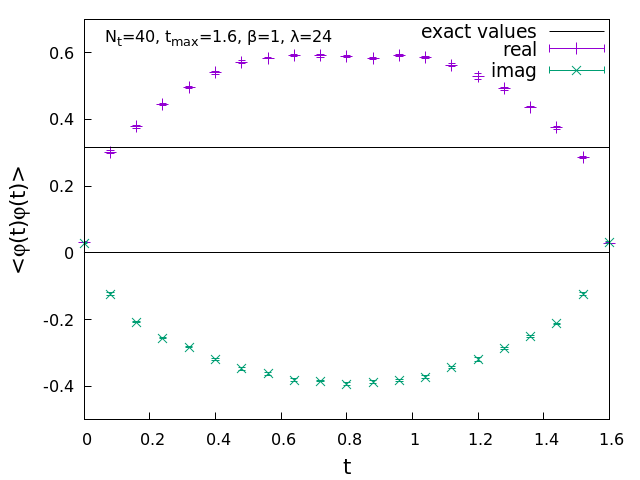}
\caption{Results for the two point function $ \langle \phi(t)^2\rangle $ for a real-time extent of 0.8 (left), 1.2 (middle) and a real-time extent of 1.6 (right), calculated with the CLE without a kernel.}
\label{goodbadfig}
\end{center}
\end{figure}

\begin{figure}[h]
\begin{center}
    \includegraphics[width=0.64\columnwidth]{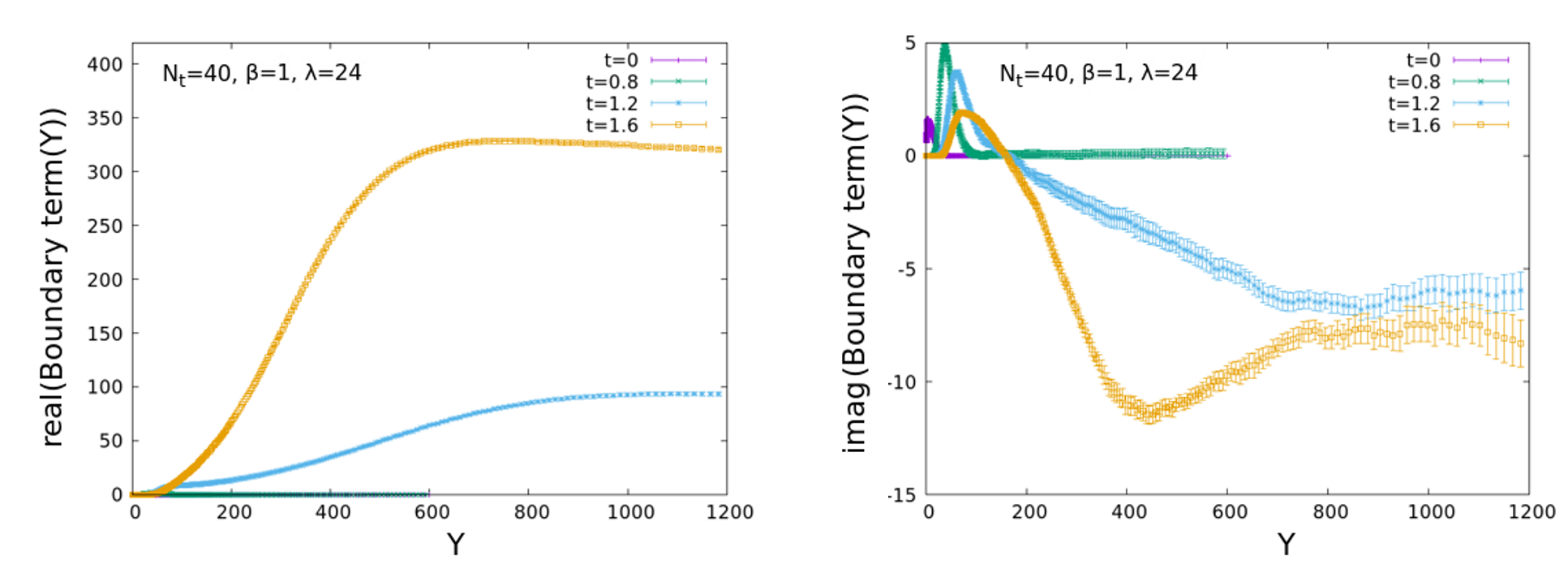}
\caption{Real part (left) and imaginary part (right) of the boundary terms for the observable $\phi^2$ for different real-time extents, as calculated from a simulation with the un-kernelled CLE.}
\label{goodbadboundary}
\end{center}
\end{figure}

\begin{figure}[h]
\begin{center}
    \includegraphics[width=0.32\columnwidth]{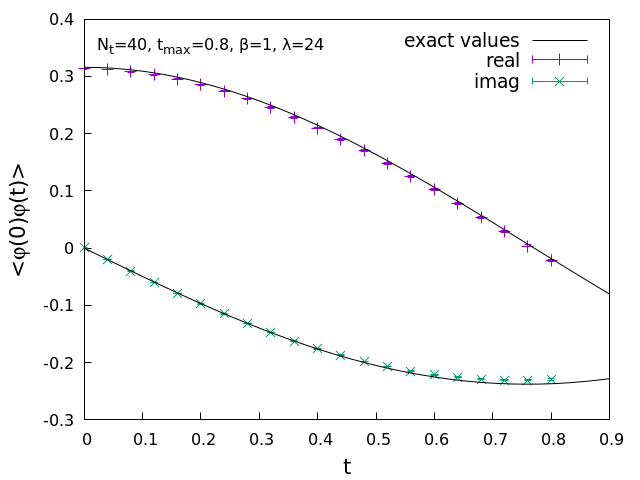}
    \includegraphics[width=0.32\columnwidth]{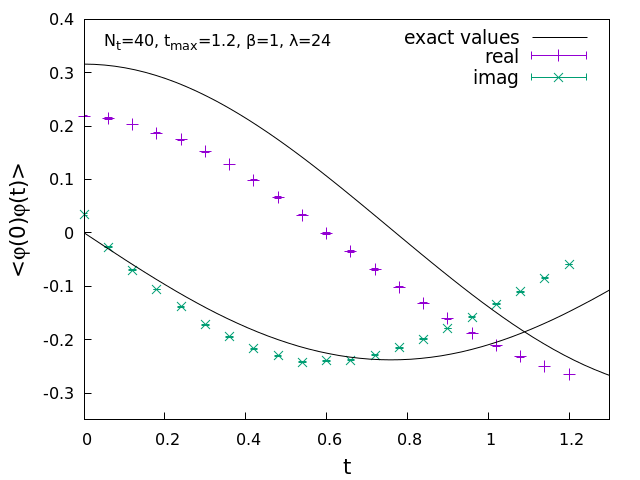}
    \includegraphics[width=0.32\columnwidth]{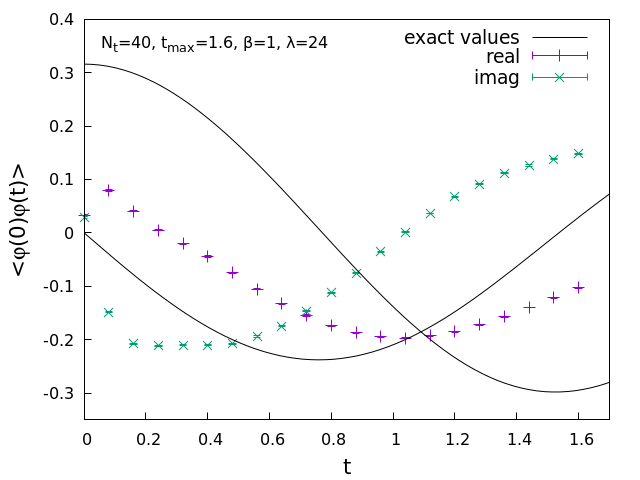}
\caption{Results for the two point function $ \langle \phi(0) \phi(t) \rangle $ for a real-time extent of 0.8 (left), 1.2 (middle) and  1.6 (right), as calculated from a simulation with the un-kernelled CLE.}
\label{goodbadfig2}
\end{center}
\end{figure}

In this subsection we show the results of the naive (un-kernelled) Langevin equation
and the boundary terms.
In Fig.~\ref{goodbadfig} we show the results for the two point function $ \langle \phi(t)^2 \rangle $
as a function of time for two different time extents. In thermal equilibrium this quantity
has to be time independent, therefore one notes that a time dependent CLE result is certainly 
wrong for the larger real-time extents. This is confirmed by comparing to the exact results gained by diagonalization,
as showed on the plot.
In Fig.~\ref{goodbadboundary} we show the corresponding boundary terms, and one observes that
they nicely signal incorrect results from the CLE dynamics itself, without needing to know the
exact results beforehand.
The unequal-time two point function $ \langle \phi(0) \phi(t) \rangle$ is shown 
in Fig.~\ref{goodbadfig2}, where, similarly to $ \langle \phi(t)^2 \rangle$, the short time extent 
shows correct results while for the larger time extents discrepancies appear.

\subsection{Introducing a constant kernel in the CLE}

\begin{figure}[h]
\begin{center}
    \includegraphics[width=0.32\columnwidth]{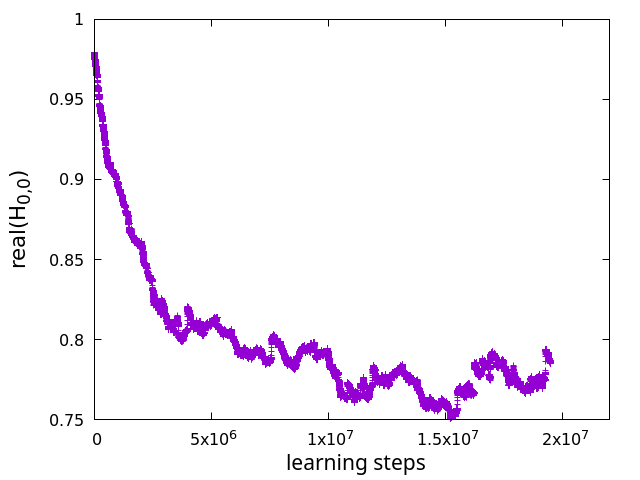}
    \includegraphics[width=0.32\columnwidth]{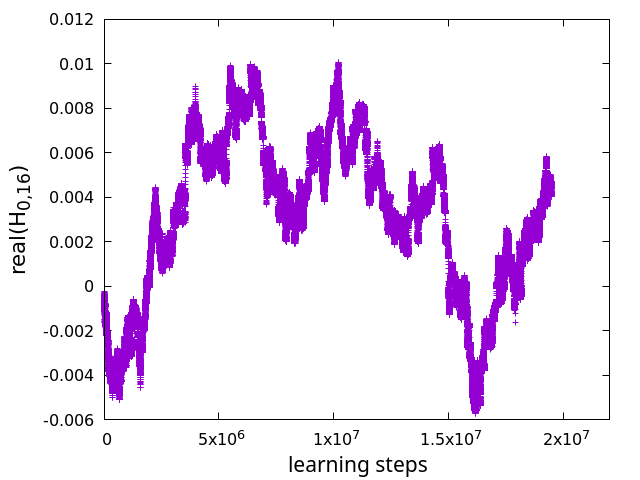}\par
    \includegraphics[width=0.32\columnwidth]{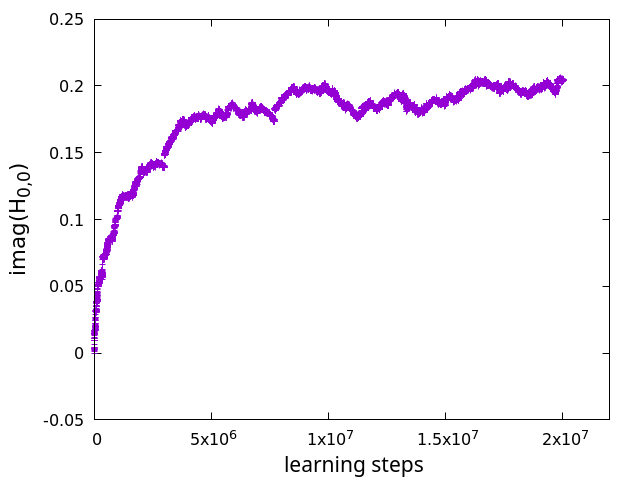}
    \includegraphics[width=0.32\columnwidth]{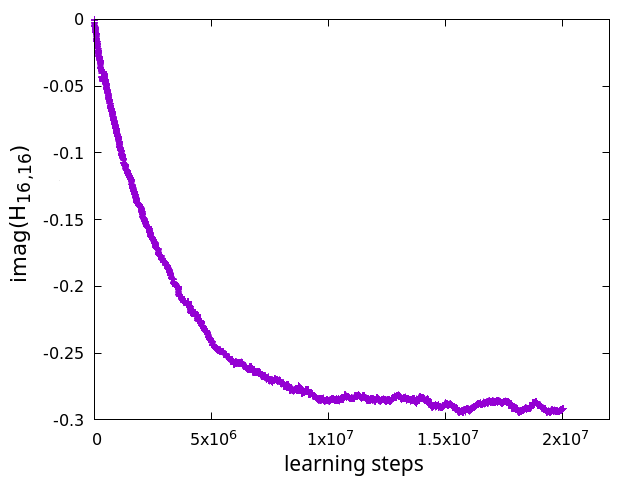}
\caption{Convergence of different matrix elements of a constant kernel for a real-time extent of 1.2 during the learning process for N=20.}
\label{kernelconvergence}
\end{center}
\end{figure}

\begin{figure}[h]
\begin{center}
    \includegraphics[width=0.32\columnwidth]{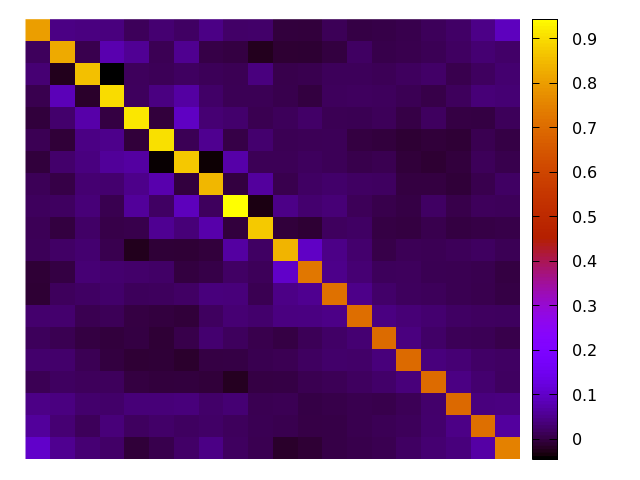}
    \includegraphics[width=0.32\columnwidth]{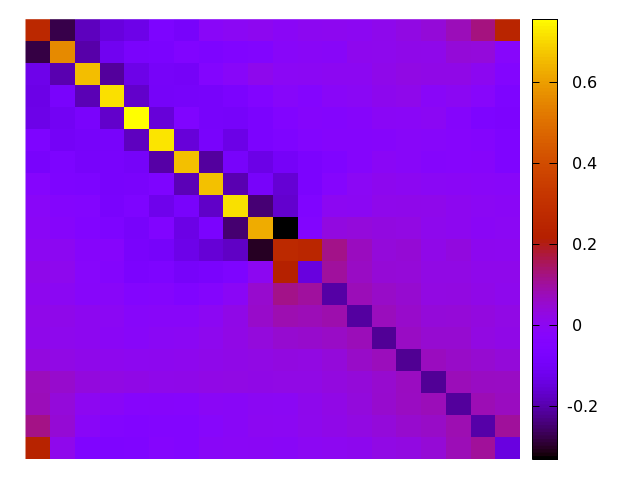} \par
    \includegraphics[width=0.32\columnwidth]{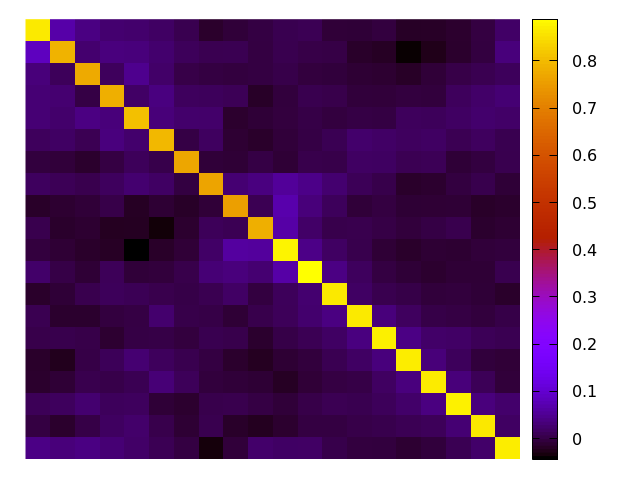}
    \includegraphics[width=0.32\columnwidth]{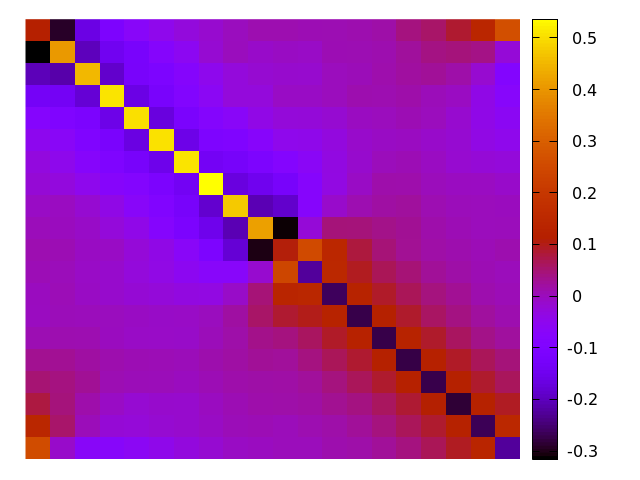}
\caption{Converged kernel matrices for real-time extents of 1.2 (real part top left, imaginary part top right) and 2.0 (real part bottom left, imaginary part bottom right),
using $N_t=20$ points on the contour.}
\label{kernels}
\end{center}
\end{figure}

In this subsection we investigate the performance of a constant kernel in the complex Langevin equation 
for the real-time anharmonic oscillator, such that the CLE is given by
\bea
\label{constkernlang}
{d \phi_i \over d\tau} = 
 -H_{ij} H^T_{jk} {  \nabla_k S }
+  H_{ij}\eta_j,
\eea
To find the optimal kernel, we use the stochastic gradient descent method.
Our loss function to be minimized is the norm 
\bea
N(\phi) = \sum_i F_1 (\textrm{Re}\, \phi)^2 +   F_2 (\textrm{Im}\,\phi)^2,
\eea
where we use $F_1=0, \ F_2=1$ below, if not otherwise noted.
The updated field from the CLE reads as 
\bea
\phi_i' = 
 \phi_i -H_{iz}(\phi) H^T_{zy}(\phi) {  \nabla_y S } \Delta \tau
+   H_{iy}(\phi) \eta_y \sqrt{ \Delta \tau}.
\eea
To make the norm minimal after the timestep, we can update our 
matrix split to real and imaginary parts $ H_{ij} = a_{ij} + i b_{ij}$
using
\bea \label{matrixupdate}
a'_{ij} = a_{ij} - \Delta L {\partial N(\phi') \over a_{ij} } , \quad
b'_{ij} = b_{ij} - \Delta L {\partial N(\phi') \over b_{ij} },
\eea
where $ \Delta L $ is the learning step size.
In practice we  average the gradient terms using the following procedure:
we start out with $ H_{ij}= \delta_{ij}$ and set the $\phi_i$ fields to zero. 
We update the Langevin equation with the current kernel, and thermalize for a Langevin time of 100.
We then collect all the configurations for a Langevin time interval of 1.
Since we use an adaptive step size with a maximal step $ \Delta \tau=10^{-5}$, this
means at least $10^5$ configurations.
We use the configurations to calculate and average the gradient terms in eq.~(\ref{matrixupdate}). We then update $a_{ij}$ and $b_{ij}$ using the averaged gradients.  After $10^6$ of such learning steps, the fields are again set to 0 (to prevent them from wondering too far) and the process is repeated until the matrix is sufficiently converged.  
We typically use $ \Delta L = 10^{-4} - 10^{-3}$. To avoid an overall rescaling of the Langevin time, we scale the matrix such that the sum squared of matrix elements 
is fixed.
In Fig.~\ref{kernelconvergence} we show a sample of the learning process for some elements of the matrix $H$, for a simulation with real-time extent 1.2 and $ N_t=20$. One observes
that the matrix elements typically equilibrate after $ 10^7$ learning steps.
The converged kernels have their largest magnitude elements on the diagonal, with significant contributions also to the first super- and sub-diagonal elements (with periodic boundary conditions). These nearest neighbour "couplings" have their largest magnitude values near the turning points of the contour. The rest of the matrix elements are small, nonzero values.
In Fig.~\ref{kernels} we show the converged kernels for real-time extents $ t= 1.2 $ and $ t=2.0$.

\begin{figure}[h]
\begin{center}
    \includegraphics[width=0.52\columnwidth]{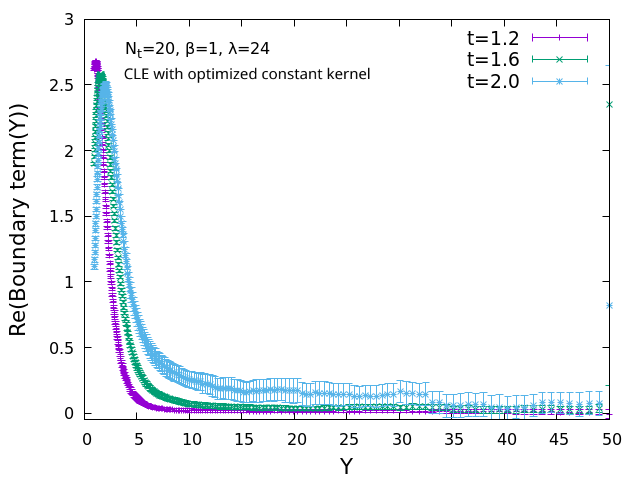}
\caption{Real part of the boundary term for real-time extents of 1.2, 1.6 and 2.0 from a kernelled CLE simulation with an optimized field-independent kernel.}
\label{boundary_kernel}
\end{center}
\end{figure}

\begin{figure}[h]
\begin{center}
    \includegraphics[width=0.32\columnwidth]{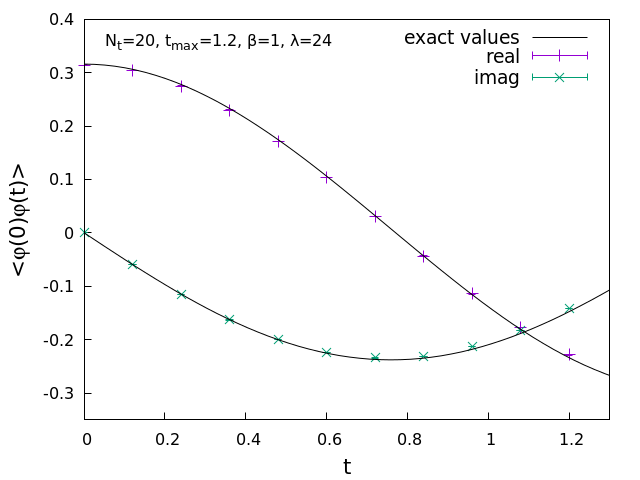}
    \includegraphics[width=0.32\columnwidth]{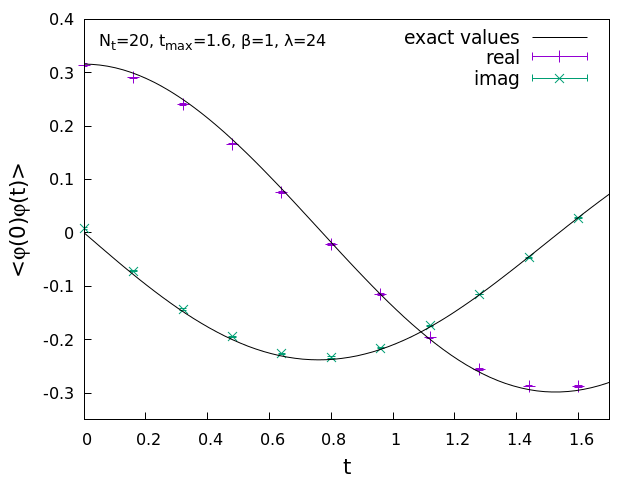}
    \includegraphics[width=0.32\columnwidth]{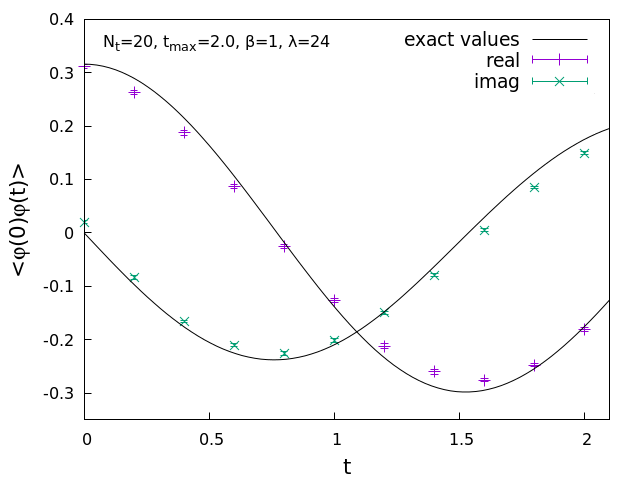}
    \includegraphics[width=0.32\columnwidth]{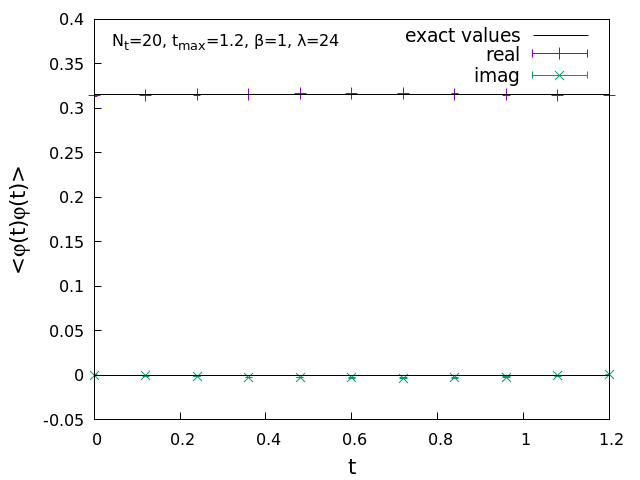}
    \includegraphics[width=0.32\columnwidth]{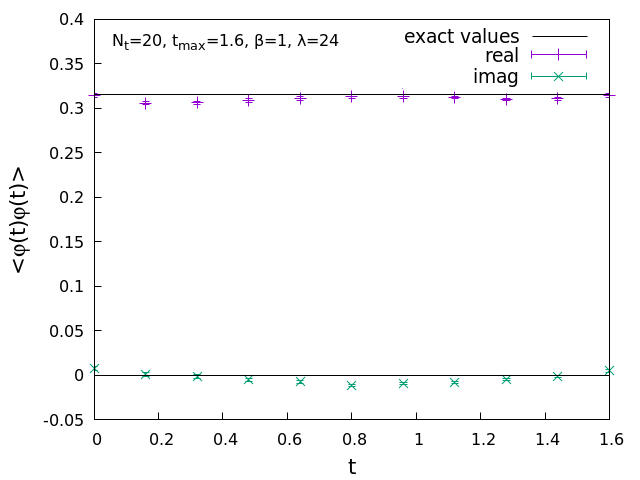}
    \includegraphics[width=0.32\columnwidth]{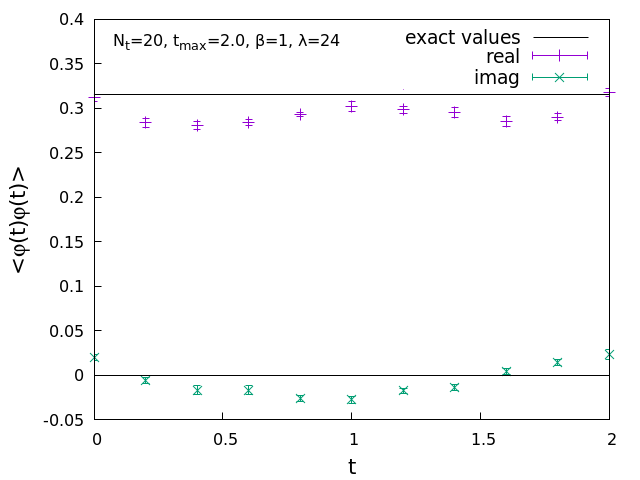}
\caption{Correlator $\langle\phi(0)\phi(t)\rangle$ (top) and $\langle\phi(t)^2\rangle$ (bottom) for optimally kernelled simulations at real time-extents of 1.2, 1.6 and 2.0, respectively. The discretisation used $N_t=20$ . }
\label{correlators_kernel}
\end{center}
\end{figure}

In Fig.~\ref{boundary_kernel} we show the boundary terms of the observable $ \langle \phi^2 \rangle $ calculated with the optimally kernelled Langevin equation. We show calculations with 
real-time extents $ t=1.2,\ 1.6,\ 2.0$. One observes that now the $t=1.2$ real-time extents shows zero boundary terms (within statistical errors). At $t=1.6$ we observe a 
small boundary term,
while the simulation
at $ t=2.0$ has the largest statistical errors, technically it is still consistent with zero, but it hints at a nonzero boundary term.
This behaviour is nicely confirmed on the correlators, shown in Fig.~\ref{correlators_kernel}. The shortest simulation at $t=1.2$ shows that results 
agree with the exact results. At $t=1.6$ small deviations from the exact results 
start to emerge, while at $t=2.0$ we see a clearly non-time translation invariant equal-time
two point function and an unequal-time two point function that deviates from the exact one.

In summary, we have shown in this section that an optimized constant kernel
can significantly extend the region of reachable real-time extents for an anharmonic scalar oscillator.

\section{Machine Learning for a field dependent kernel}
\label{secml}

In this section we study field dependent kernels. To keep complications to 
a minimum, first we study a simple one variable toy model with the action
\bea
S(x) =  {1\over 2 } \sigma x^2 + {1 \over 4 } \lambda x^4.
\eea
This model was previously investigated in \cite{Okano:1991tz}.
Here we test the model using the parameters $ \sigma=-1 +4 i, \ \lambda=2$,
where we have the approximate result $ \langle x^2 \rangle = 0.106875 -i 0.440711$, as calculated by e.g. numerical integration using the trapezoid rule.

To overcome the problem of wrong convergence we introduce a field 
dependent kernel $K(z)$, such that the CLE is modified to eq. (\ref{singleKCLE}).
We use various kernels: first, we use the ansatz
\bea
K(z) = C_1 e^{-z^2/a} e^{-i\Theta_1} + C_2 \left(
1-e^{-z^2/a} \right) e^{-i \Theta_2}
\eea
inspired by \cite{Okano:1991tz}, where parameters $ C_1,\, C_2,\, \Theta_1,\, \Theta_2,\, a$ are to be optimized. (In \cite{Okano:1991tz} they were chosen explicitly using theoretical considerations.)  
We use the loss function $ N = F_1(\textrm{Re} z)^2 +F_2(\textrm{Im} z)^2 $, and a similar 
optimization procedure to the one used in section \ref{realtimesection}.
In some cases the system achieves a small loss function by simply choosing a small kernel 
(that is, small $C_1$ and $C_2$). This is equivalent to using very small Langevin
timesteps, which means the field doesn't have time to grow large and reach larger 
loss function values. As this is an undesired behavior,
we introduced a new term in the loss function which punishes small kernel values.
In practice one can use e.g. this expression
\bea
L_\textrm{timestep} =  { 0.01 \over \sum_{ij} (\textrm{Re} H_{ij})^2
  + ( \textrm{Im} H_{ij} )^2 },
\eea
where we used the notation for a general matrix kernel. With this prescription 
the optimization procedure works well and finds a kernel such that the kernelled CLE 
delivers results close to the exact results.

Second, we use a neural network to build
a $K(z)$ function. The first difficulty that needs to be addressed is the choice of the
activation function, as we want to keep a holomorphic kernel. A kernel with some 
singularities is expected to be acceptable, as long as the process vanishes sufficiently 
fast near the singularities. A general non-holomorphic kernel, dependent separately on the real and imaginary parts of the field is not expected to work. 
If one wants to have no singularities on the complex plane (except at infinity), on is basically left
with polynomials, exponentials, and combinations (sums or products) thereof.
For the purposes of this study, we have chosen the activation function
\bea
y(z) = e^{ -z^2 /a }.
\eea
Note that the universal approximation theorem for complex neural networks
does not hold for holomorphic functions \cite{voigtlaender2023universal}. 
We also did some experiments with the activation function
\bea
y(z) = {1\over 2} \left( 1 + \cos( \arg (z) ) \right) z,
\eea
which violates holomorphicity, but the universal approximation theorem is valid
\cite{voigtlaender2023universal}.
Similarly to the previous case, we need a term in the loss function which does
not allow small kernel values.
With an optimization procedure similar to what has been used before, we get results
close to the exact ones, however in the case of the non-holomorphic kernel we 
see significant deviations especially in the imaginary part  
of the observable $ \langle z^2 \rangle$.

We also investigated field dependent kernels for the 
real time evolution studied in Section \ref{realtimesection}.
The kernels used were of the form 
$ H (\phi) = H_\textrm{const} + H_\textrm{nn} (\phi)$, where 
$H_\textrm{const} $ is a constant kernel we have first optimized 
using the procedure in Section \ref{realtimesection}. $H_{nn}(\phi)$
is a field dependent kernel that is given by a neural network.
We have chosen again the activation function $ \exp(-z^2 ) $.
The neural network consisted of 4 layers with 32,32,64,and $N_t^2$ nodes, such that
its output can be used as an $N_t$ by $N_t$ complex matrix. 
Using this formulation we can efficiently compute the kernel 
as well as its derivative (which we need for the CLE), with forward and backward propagation
on the network.
We choose random
initial values for the network parameters such that $H_\textrm{nn}$ are small.

\begin{figure}[h]
\begin{center}
    \includegraphics[width=0.32\columnwidth]{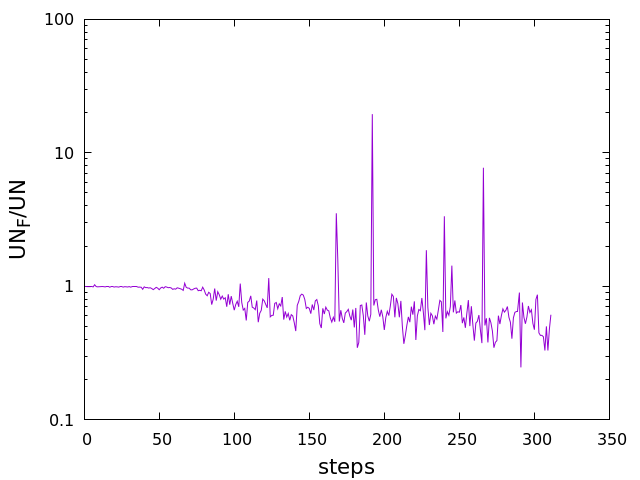}
\caption{Imaginary norm with a field dependent kernel UN$\mathrm{_F}$ divided by the imaginary norm with a field independent kernel UN over optimization steps.}
\label{UN_spikes}
\end{center}
\end{figure}

In Fig.~\ref{UN_spikes} we show the behavior of the imaginary norm $ N(\phi) = \sum_{x,t} (\textrm{Im} \phi_t(x))^2 $ in ratio to the average imaginary norm with a constant kernel, 
as the training of the network progresses. 
One notes that the majority of the values are indeed 
smaller for the field dependent kernel, however there are huge spikes in the values.
This is probably caused by the neural network amplifying the tails 
of the complexified distributions in the imaginary directions, as $ \exp( (\textrm{Im} z)^2 )$ is growing rapidly. Unfortunately this also means that expectation values 
of observables are contaminated by the spikes, and thus are unusable.
In summary, field dependent kernels for the real time oscillator are showing 
promising first results, but more work is definitely needed to solve its
problems.

\section{Conclusions}
\label{secconc}
In this paper we have investigated the application of kernels to
solve the problem of the Complex Langevin approach: occasional 
convergence to incorrect results due to 
boundary terms or wrong spectral properties of the $L_c$ operator \cite{Seiler:2023kes}.

We studied an anharmonic quantum oscillator in thermal equilibrium on a 
Schwinger-Keldysh-like complex time-contour. The naive CLE gives correct results
at small real-time extents and wrong results at large real-time extents, as confirmed by measurement
of the boundary terms, as well as comparison to exact results (calculated by diagonalization of the 
Hamiltonian). The incorrect results are also signalled by a time dependent equal-time two point 
function. 

Next, we have introduced a constant matrix kernel in the CLE. We have optimized the kernel
using stochastic gradient descent by requiring that the distance of the complexified 
distribution from the real manifold is minimal. This in turn has led to sizeable 
reduction of the boundary terms. The reachable real-time extent
(where the boundary terms are zero within statistical errors) with the optimized kernel 
is roughly twice as large as with the naive CLE, for the parameters used in this study.

Finally we have explored possibilities of field dependent kernels, for which we use 
a simple ansatz or we 
represent them with a neural network. As we want to choose a holomorphic kernel function, this requires 
a careful choice of the neural network architecture and especially the activation function.
We have studied a one variable toy model and we have seen that the optimization
procedure used before works also in this case, if the loss function is modified 
such that low kernel values are avoided (these would correspond to "slowing" the Langevin
dynamics such that field values remain stuck).
Promising first results were also gained for the field dependent kernel 
of the quantum oscillator, however the results we gained tend to have a long tailed distribution,
probably caused by large magnitude kernel values, so more work is needed to better understand this behavior, 
perhaps a different network architecture would be more suitable for the complex kernels. We plan to come back to this issue in a subsequent study.

Note that an independent but similar study has appeared recently \cite{Alvestad:2022abf},
with similar results and some important differences:
The loss function in \cite{Alvestad:2022abf} used a condition on the drift terms to 
achieve field-independent kernels such that the drift attracts field values to the origin, and also used boundary terms, symmetries of the system and results from euclidean simulations in the loss function. 
In contrast, here 
we use a simpler loss function which just aims at the minimization of the 
average distance of fields from the real axis (or the origin), and we achieved robust convergence 
to an optimal kernel. 
In this study we also report on our first experiments with field dependent kernels.

One of the main advantages of the CLE approach to real-time evolution is the volume scaling of the 
costs of the simulations, as CLE is expected to scale roughly linearly with the volume, in contrast with
the exponential scaling for diagonalization. This makes studying field theories feasible, and an 
extension of the study of scalar field theory with nonzero spatial extent is already underway \cite{arsinprep}. 

In summary, we have shown that using kernels can decrease boundary terms such that 
the CLE delivers (within statistical errors) correct results. In particular for the 
scalar oscillator, an optimized constant kernel allows calculation of the observables on real-time contours that are significantly longer than what the un-kernelled CLE allows.
First indications show that field dependent kernels might lead to an even 
better behavior, however, more work is needed in this line of study.

\begin{acknowledgments}
D.~S.~ acknowledges the support of the Austrian Science Fund (FWF) through the Stand alone Project P36875.
The numerical simulations for this project were carried out on GSC, the 
computing cluster of the University of Graz.
\end{acknowledgments}


\bibliography{mybib}

\end{document}